# Charge Crowding in Graphene-Silicon Diodes

Muhammad Abid Anwar, Munir Ali, Dong Pu, Srikrishna Chanakya Bodepudi, Xinyu Zhu, Xin Pan, Jianhang Lv, Khurram Shehzad, Xiaochen Wang, Ali Imran, Yuda Zhao, Shurong Dong, Yang Xu, Senior Member, IEEE
, Bin Yu, Fellow, IEEE, and Huan Hu

*ABSTRACT*-The performance of nanoscale electronic devices based on a two-three dimensional (2D-3D) interface is significantly affected by the electrical contacts that interconnect these materials with external circuitry. This work investigates charge transport effects at the 2D-3D ohmic contact coupled with the thermionic injection model for graphene/Si Schottky junction. Here, we focus on the intrinsic properties of graphene-metal contacts, paying particular attention to the nature of the contact failure mechanism under high electrical stress. According to our findings, severe current crowding (CC) effects in highly conductive electrical contact significantly affect device failure that can be reduced by spatially varying the contact properties and geometry. The impact of electrical breakdown on material degradation is systematically analyzed by atomic force, Raman, scanning electron, and energy dispersive X-ray spectroscopies. Our devices withstand high electrostatic discharge spikes over a longer period, manifesting high robustness and operational stability. This research paves the way towards a highly robust and reliable graphene/Si heterostructure in futuristic on-chip integration in dynamic switching. The methods we employed here can be extended for other nanoscale electronic devices based on 2D-3D interfaces.

## I. INTRODUCTION

Vast multidisciplinary applications of graphene have promoted enormous developments and made breakthroughs in the high-tech fields. The extraordinary potential of graphene lies in its remarkable thermal conductivity [1] dangling bond-free interfaces, efficient charge transfer [2] high electron mobility [3], [4] mechanical resilience [5], [6] high-temperature sustainability [7] and broadband optical absorption and sensing [8], [9] which make it an ideal material for high-performance electronics and optoelectronics.

However, the main physical limitation of graphene lies in the weak interface contact between graphene and metal electrode [10] which results in excessive heat generation, the creation of local hot spots [11] electromigrations [12] and consequently compromising the lifetime and reliability of the device [13]. One of the critical factors limiting the performance, robustness, and durability of electronic devices through electrical conduction is the current crowding (CC) effect, defined as an inhomogeneous distribution of current density at two dissimilar highly conductive contacts [10], [14]–[24]. Numerous researches have demonstrated the CC effects in 2D materials, e.g., $MoS_2$ or graphene [25]–[27] with different strategies to reduce the CC effect and improve the current transport in the electrical contact [28]–[30]. Understanding the 2D-3D contact failure mechanism is critical for the real future application of 2D-3D heterostructure devices.

The 2D-3D Schottky junctions require special attention to understand how they differ from traditional metal-semiconductor Schottky and the key benefit of using a 2D heterostructure (graphene) as an alternative of metal. Typically, metal-based electronic devices suffer from high noise, Fermi level pinning, non-ideal interfaces, large surface, and trapping sites limiting efficient charge transfer between metal and semiconductor [31], [32]. The Gr/Si Schottky junction has been reported as fast and highly responsive photodiodes with broadband and high sensing applications [33], [34]. However, the impact of inherent time-dependent instabilities in the charge transport at the 2D-3D interface (graphene-metal/$SiO_2$/Si) demands special consideration. The weak 2D-3D contacts (graphene-metal) [35] remarkably deteriorate the performance of graphene-based electronic devices. Graphene is highly sensitive to the environment due to its 2D nature and is strongly influenced by metal contacts. The electrical properties of graphene can be tuned when in contact with different metals such as Pt, Au, and Cu [36]. The charge loses a fraction of energy as phonon vibrations, causing scattering in the lattice and raising the material temperature [37]. The non-uniform distribution of the carriers across the graphene-metal contacts forms thermal hot spots, deteriorating device performance and causing material failure [38].

This work is supported by the National Natural Science Foundation of China (NSFC) (Grant No. 92164106, 62090030, 62090034, 62104214) and the Fundamental Research Funds for the Central Universities (2021FZZX001-17). We thank Prof. Wenchao Chen, Dr. P. Pham, Dr. Wei Liu, Dr. Hongwei Guo, Dr. Lingfei Li, Dr. K. Dianey, Xinyu Liu, Xiaoxue Cao, Muhammad Malik, Feng Tian for fruitful discussion and valuable comments on the manuscript. Corresponding author: Huan Hu, Bin Yu and Yang Xu). Muhammad Abid Anwar, Munir Ali, Dong Pu, Srikrishna Chanakya Bodepudi, Xinyu Zhu, Xin Pan[1], Jianhang Lv, Khurram Shehzad, Xiaochen Wang, Ali Imran, Yuda Zhao, Shurong Dong, Huan Hu, Bin Yu, and Yang Xu, are with School of Micro-Nano Electronics, ZJU-Hangzhou Global Scientific and Technological Innovation Center, ZJU-UIUC Joint Institute, State Key Laboratory of Silicon Materials, Zhejiang University, Hangzhou, 310027, China.
(E-mail: *yangxu-isee@zju.edu.cn, yu-bin@zju.edu.cn, huanhu@intl.zju.edu.cn*)

## II. EXPERIMENTAL METHOD AND MEASUREMENTS

The Gr/Si Schottky devices are fabricated on a low doped n-type Si wafer with a 300 nm, 100 nm and 18 nm thick $SiO_2$ layer. Au/Ti (70/10, 60/10 and 40/10 nm) electrodes are patterned by the 1st lithography followed by e-beam evaporation and lift-off. In the 2nd lithography, the Si window is exposed, and the wafer is immersed into buffered oxide etchant to remove the oxide layer. Then, graphene is chemically transferred onto the Si window to form a Schottky junction. Finally, the 3rd lithography and $O_2$ plasma remove residues on the surface. A metallic Ohmic contact is formed on the backside of the Si substrate. We used a TLP tester (Barth pulse curve tracer semiconductor analyzer, model 4002) with short ESD pulses (100 ns pulse widths and 10 ns rise times) for the Electrostatic testing. SEM and XRD images were obtained using a Hitachi S4800 field emission microscope at an acceleration voltage of 5 kV. A confocal Raman microscope (Senterra, Bruker) and a Renishaw Invia with a 532 nm excitation wavelength were used to measure Raman spectroscopy.

## III. RESULTS AND DISCUSSION

This work employs Gr/Si Schottky junction to signify the charge transport at the 2D-3D contact under high electrical stress. The 3D schematic of the Gr/Si Schottky junction is presented in Fig. 1(a). The surface roughness model consists of four regions, graphene-metal lateral and edge contact ($R_{G-M}$), suspended graphene region ($R_G$-$SiO_2$), and Gr/Si ($R_G$-$Si$) interface to measure the surface roughness shown in Fig. 1(b). The low density of state (DOS) close to the Dirac point limits the charge injection from metal to graphene. The amount of charge transport across the metal-graphene interface is controlled by the work functions and contact resistance presented in Fig. 1(c). The current-voltage characteristics of the Gr/Si Schottky diode showed rectification behavior with ideality factor (η = 1.1) and series resistance ($R_s$ = 480 Ω) calculated by Cheung's method (Fig. 1(d)) [39]. Reducing the contact roughness to minimize the contact resistance will improve the device charge transport and reliability. The influence of contact, surface morphology and roughness of the 2D-3D interface of the intrinsic Gr/Si junction is exhibited through AFM and SEM, as shown in Fig. 1(e) and Fig. 1(f). The AFM images with the different roughness of the junctions between dissimilar materials composed of roughness at graphene-metal ($R_{G-M}$) = ~4 nm (lateral), ~35 nm (edge site), $RG$-$SiO_2$ = ~10 nm, and $R_{G-Si}$ = ~3 nm for 18 nm thick oxide layer.

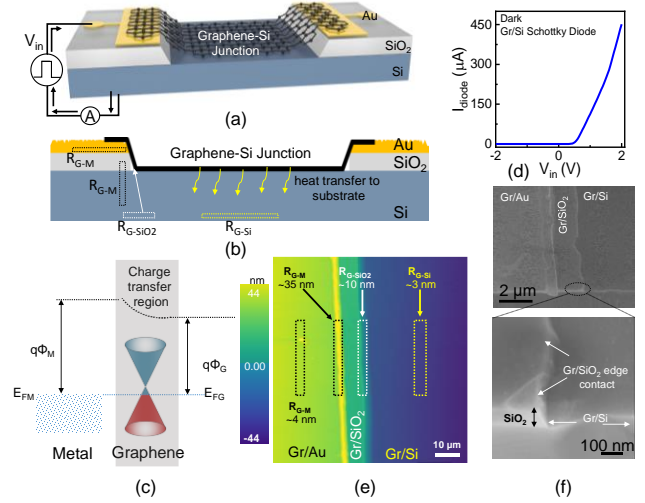

**Fig. 1.** (a) 3D schematic of CVD Gr/Si Schottky diode. (b) The cross-section image shows the effect of charge injection at graphene-metal, Gr-$SiO_2$ contact (insufficient for a heat sink) and Gr/Si (heat transfer to the substrate). (c) Energy band diagram of metal–graphene contact (The Fermi levels are aligned and show a charge transfer mechanism at the graphene-metal interface). (d) Static I-V characteristic (e) AFM morphological between dissimilar materials Gr/Au (lateral and edge), Gr/$SiO_2$ (edge), and Gr/Si are showing the surface roughness at four different sites. (f) SEM image of the device structure.

## IV. ELECTROSTATIC DISCHARGE (ESD) FAILURE MECHANISMS OF GRAPHENE-SI DIODE

The 2D-3D contact, like graphene-metal contact, suffers from interfacial defects and weak contact, compromising the current carrying capability and switching speed of the device. When the excessive heat energy is not dissipated properly, it can lead to the device breakdown that creates permanent damage to the material [38], [40]–[43].

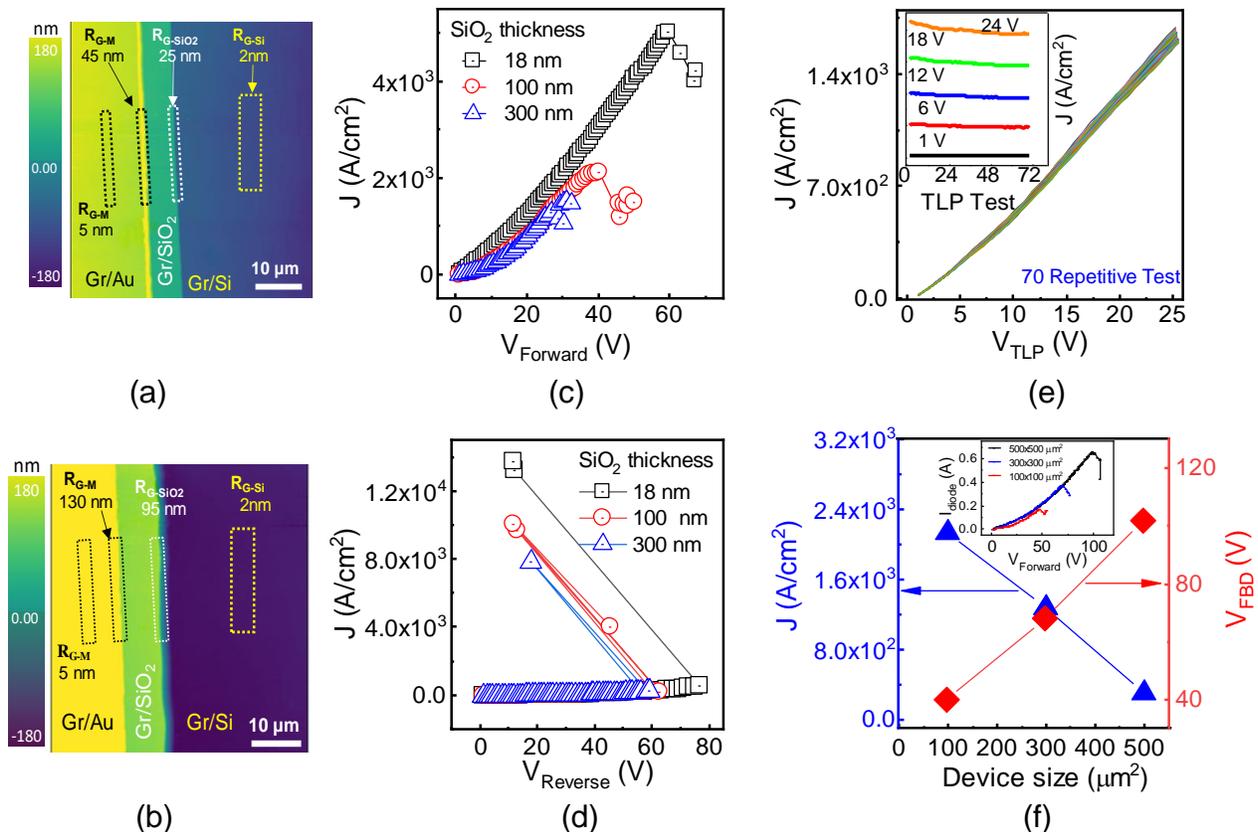

**Fig. 2.** AFM surface morphologies and roughness (a) 100 nm (b) 300 nm oxide thickness. TLP testing shows (c) forward breakdown (d) reverse breakdown characteristics for Gr/Si Schottky diode with different $SiO_2$ layers. (e) The device stability test exhibited 70 repetitive measurements regarding increasing $V_{TLP}$ stress (for 300 nm oxide layer). (f) Relationship of current density and breakdown voltages with varying the device sizes regarding a gradual increment of $V_{TLP}$ stress in forwarding mode (inset: shows breakdown current-voltage characterization of different window size Gr/Si Schottky diode)

A part of the excess energy gained by electrons from an external source (e.g., an electrical bias) is transferred to the lattice via phonon emission, increasing the lattice temperature due to Joule heating. Therefore, heat mitigation is an essential factor that significantly impacts the graphene-based structure, where 2D materials demand extensive heat sink contact with the underlying substrate. Here, we experimentally investigated the influence of Joule heating that is caused by the CC effect in 2D-3D contacts. AFM morphological roughness measurements at graphene-metal (lateral and edge site), Gr/$SiO_2$ and Gr/Si for 100 and 300 nm thick oxide layer are presented in Fig. 2(a) and Fig. 2(b). The AFM result reveals the prevailing roughness distribution at the edge contacts. It confirms that lowering these thicknesses minimize interface gaps, reduce the roughness of edge contacts and improve the charge transport. Moreover, we perform thermal annealing to smooth the gold surface as a result shows low edge roughness, which significantly improves thermal mitigation. The larger gaps at the interfaces also stretch the graphene laid on the contact and substrate, producing more wrinkles enhance phonon vibration resulting in earlier device failure for the thicker oxide layer.

We executed a TLP testing with a sharp rising time (10 ns) and pulse width (100 ns). The TLP test is suitable for ensuring device reliability and stability under ESD stress [44]. The excess heat caused by graphene electrons cannot be dissipated effectively through the thicker oxide resulting in a temperature rise and enhanced electron-phonon scattering. A thick oxide layer creates a large area of suspended graphene that has no direct contact with the solid surface, which serves as an inefficient heat sink. Therefore, suspended graphene is easy to heat up to high temperature under high electrical stress and gets damaged earlier than graphene with contact with a solid substrate. As a result, the transferred heat can quickly spread and provide a better heat sink, suggesting efficient heat mitigation for thinner oxide [45], [46].

The TLP results verify that a thinner oxide layer reduces surface roughness and contact resistance, consequently increasing the graphene-substrate contact area to improve the electron transportation and heat mitigation, resulting in extended device breakdown voltage and significantly enhance the current density, as shown in Fig. 2 (c). Intrinsically, the Gr/Si Schottky junction possesses a low dark current in reverse mode and a high on/off ratio, applicability as a fast Gr/Si Schottky diode.

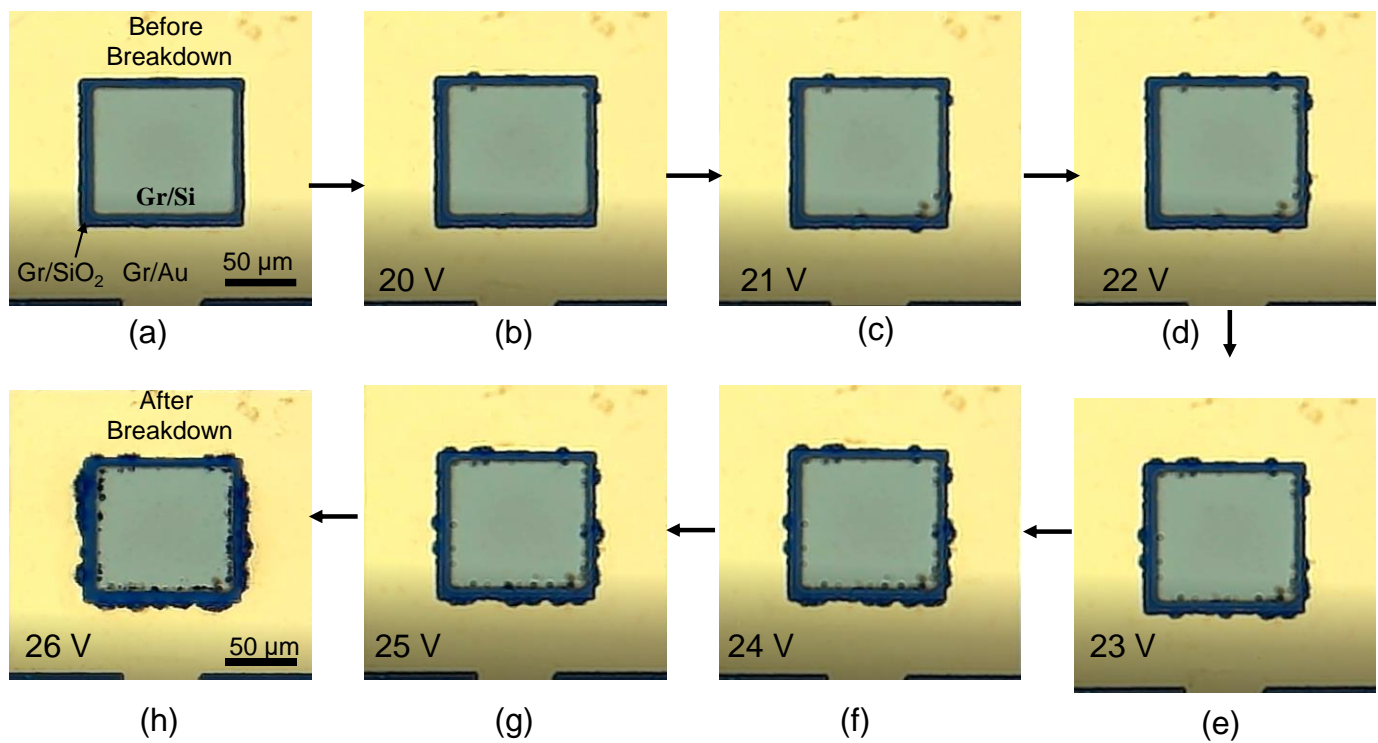

**Fig. 3.** A transition from soft to hard breakdown with increasing voltage is visible through optical microscopy images of Gr/Si Schottky devices regarding a sequence of device stressing under TLP pulses before (a) and after (b-h) electrical stress with rising times (10 ns) and pulse widths (100 ns) for 300 nm oxide layer.

The reverse bias TLP testing justifies the low leakage current resulting in large breakdown voltage, demonstrating the device ability to bear high-amplitude ESD spikes in Fig. 2(d). we performed a repetitive TLP test (70 times) on the same device to investigate the repeatability and explore any evident changes in the I-V characteristic due to electrical stress. As a result, the I-V characteristic did not deviate, demonstrating Gr/Si junction is robust enough (Fig. 2(e)). The possible charge movement, CC effect, the high current density at 2D-3D contact, and their impact on the time-dependent dynamics are directly dependent upon the junction area of the device. A device with a smaller junction exhibits lower breakdown voltages than that with a larger junction due to the high current density. In Fig. 2(f) increasing the Gr/Si junction area from $0.1 \times 0.1$, $0.3 \times 0.3$, to $0.5 \times 0.5$ mm$^2$ significantly pushing the high forward breakdown voltage ($V_{BD}$) as illustrated in Fig. 2(f) inset.

The excessive charge movement across the junction introduces a heating effect along the graphene channel. It is observed that the heat is appropriately mitigated, where graphene is laterally contacted with metal and substrate. The graphene near the edge contacts appears to be more resistive, less efficient at managing heat, and prone to device failure. The stress accumulation and defect generation are more severe at high ESD, which produces hot spots and localized heating at contact corners due to field crowding [47]. Furthermore, as electrical stress increases, cracks/defects grow further, melting the metal electrode and causing device breakdown (from soft to hard), as presented in Fig. 3(b-g). Fig. 3(h) shows a complete breakdown at 26 V, with no conductive path available for graphene on Si to contact the Au electrode for the further utility of the Gr/Si junction. By carefully monitoring the visuals, the failure region was detected at Gr/Au and Gr/SiO$_2$ edges but had a minor impact on the Gr/Si junction.

Furthermore, we performed the SEM of three different functional regions of the device to determine the exact graphene degradation sites leading to device failure, as presented in Fig. 4(a) and Fig. 4(b). Raman spectroscopy is an appropriate characterization tool for evaluating the atomic disorder of the material[48]–[51]. We spectroscopically measure 10 spots that encompass the Gr/Au (1,2), Gr/SiO$_2$ (3-6), and Gr/Si (7-10) to probe the material degradation under high electrical stress. In Fig. 4(a) at position (1-10), the Raman spectra demonstrated an excellent sp$^2$ hybridized monolayer graphene with high ($I_{2D}/I_G > 2$) and low ($I_D/I_G < 1$) peak ratios. The D band corresponds to scattering from local defects or disorders present in carbon, and the G band originates from the in-plane tangential stretching of the C−C bonds in the graphitic structure [52].

The catastrophic effect of electrical breakdown occurs under high electrical stress, demonstrated in Fig. 4(b). Our result indicates that failure occurs due to excessive phonon vibrations resulting in C-C bond breaking that finally electrically disconnects the graphene in the device. The Raman spectra of the areas from 3 to 6 conclude that extensive current activity boosted the edge contact heating effects, leading to material failure. At the positions (7,8) from the SEM in Fig. 4(b), the 2D peak appears at 2690 cm$^{-1}$ with an emerging D' adjacent to a high D peak at (1620/1350 cm$^{-1}$), owing to graphene self-roll (folded) at after covalent bond breaking [53]. In contrast, graphene retains its continuity and symmetry at the Gr/Si junction, as represented by positions 9 and 10.

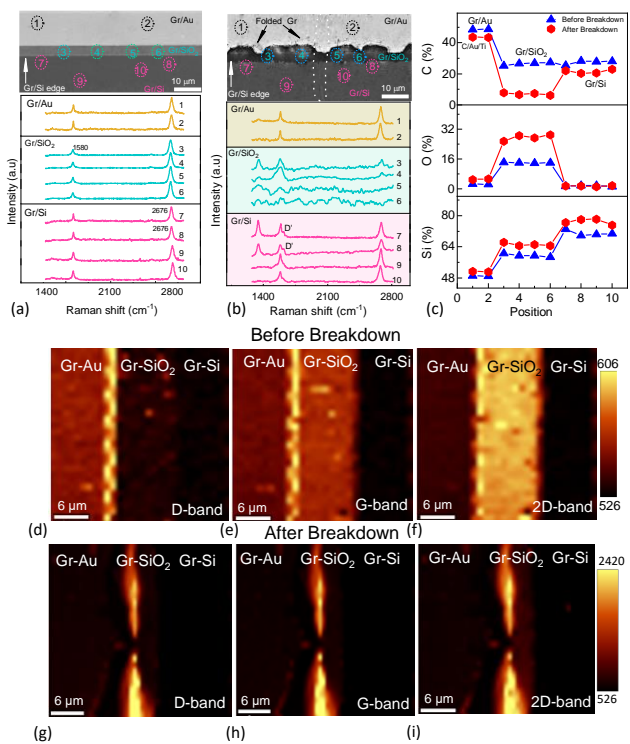

**Fig. 4.** SEM images and Raman spectra of Gr/Au, Gr/SiO$_2$ and Gr/Si contacts (a) before and (b) after breakdown, respectively. The D and D′ peaks, indicative of disorder, increase and become more prominent with increased electrical stress. Conversely, the D, G, and 2D peaks weaken, and graphene ruptures at the edge contact (points 3-6), reducing carbon atoms after an electrical breakdown. (c) EDX data of C/ Au/Ti, O, and Si elements in (a) and (b). (Note: SEM images in (a) and (b) with point distributions across the three regions of Gr/Au (positions of 1-2), Gr/SiO$_2$ (positions of 3-6), and Gr/Si (positions of 7-10). (d-i) Raman mapping at ~1340 cm$^{-1}$ (D-band), ~1580 cm$^{-1}$ (G-band), ~2680 cm$^{-1}$ (2D-band) before and after breakdown.

In Fig. 4(d-f), Raman mapping shows significantly high defect density at the edge contact, enhancing the charge crowding, excessive phonon vibrations at high current density, and C-C bond breaking at Gr/Au and Gr/SiO$_2$ edge contact (Fig. 4 (g-i)). Finally, the energy dispersive X-ray (EDX) spectroscopy approach is used for qualitative materials analysis with the elemental composition at points 1-10 of the device before and after the breakdown in Fig. 4(a) and Fig. 4(b). In Fig. 4(c), the elemental compositions of all associated components in the device, such as carbon C, oxygen O, and Si, are shown in percentages versus points 1-10. The graphene was ruptured at the edge contact (points 3-6), reducing carbon atoms, as evidenced by EDX spot mapping. The decrease of C and increase of O percentages correspond to C-C covalent bond breaking at Gr/SiO$_2$ areas.

Moreover, we examined the role of graphene nanoribbons (GNRs) in interaction with a substrate under high electrical stress. The GNRs exhibits remarkable electrical and optical properties, and numerous studies highlight the innovative characteristic of advanced nanotechnology [54]–[57]. Fig. 5(a) and Fig. 5(b) demonstrates static IV characterization and pre-failure SEM images of GNRs device. Fig.5(c) shows the Raman spectra of pristine GNRs, with a low D-peak at 1340 cm$^{-1}$, G peak at 1580 cm$^{-1}$, and a 2D peak at 2680 cm$^{-1}$ as the $I_{2D}/I_G>2$ and $I_D/I_G<1$ peak intensity ratios. The D peak is not observable in pristine graphene because of crystal symmetries [58]. Several studies have demonstrated different densities of defects in graphene ribbons, including symmetry-breaking, disorder, crystal alignment, lattice, and edge defects that can limit the device performance [55], [59]–[66].

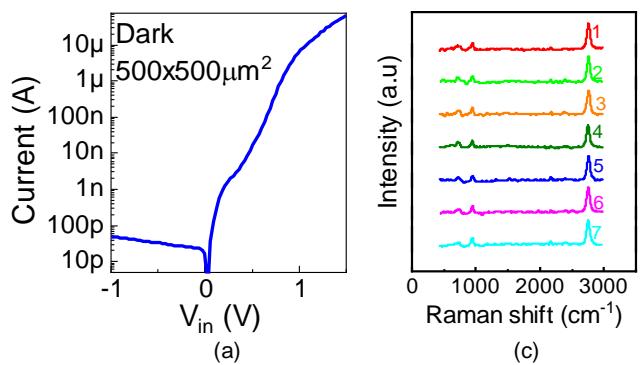

**Fig. 5.** (a) Static IV characterization. (b) SEM images. (c) Raman spectroscopy of Graphene-Silicon Schottky ribbon device at different points.

However, there has been limited research into how the density of defects influences GNRs properties under high electrical stress. The electrical charge injection generates thermal heat and produces phonon scatters, significantly affecting the Gr/Si junction. Raman spectroscopy uses to quantitatively identify the defects in the graphene ribbons by observing D, G, 2D, D′, and $I_D/I_G$ peak intensity ratios.

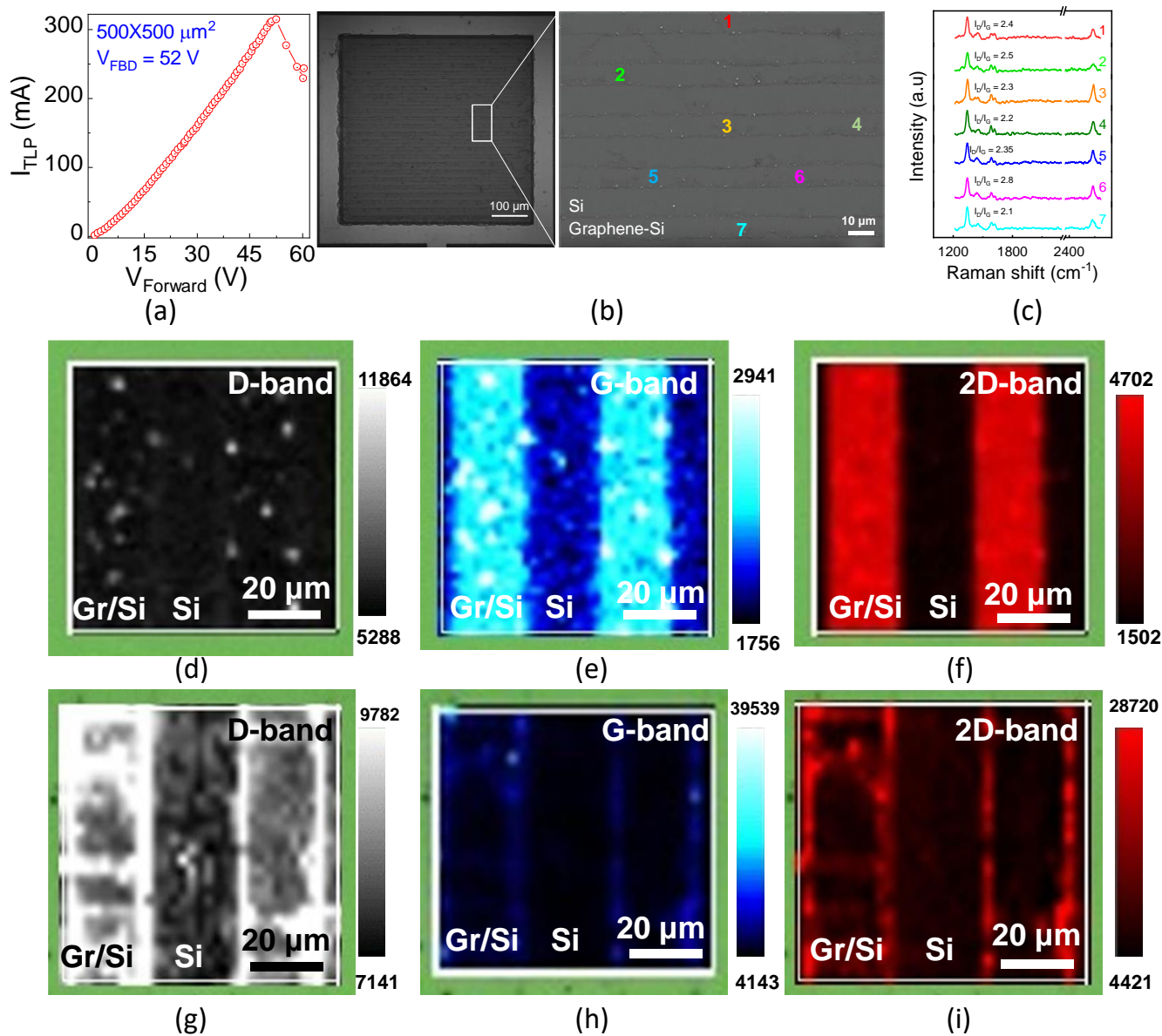

**Fig. 6.** (a) TLP forward breakdown of Gr/Si Schottky diode with GNR (b), (c) Post-failure S.E.M. images show that higher defect density creates more phonon dispersion after electrical stress. (d) The evolution of D, G, and 2D peaks was observed, and Raman peak intensity ratios of the D, D′, and $I_D/I_G$ ratio increased after an electrical breakdown. Note: The higher D, D′, and $I_D/I_G$ indicate the formation of more defects in the graphene ribbons. Raman mapping of a G.N.R./Si Schottky junction. Raman map at (e) ~1340 cm$^{-1}$ (D-band), (f) ~1580 cm$^{-1}$ (G-band), (g) ~2680 cm$^{-1}$ (2D-band) before and after a breakdown. (e-f) Pre-failure G.N.R. had sharp and distinct uniform boundaries. (h-j) Sharp GNR changed into non-uniform and irregular termination at edges resulting from the disruption of GNR due to electrical stress.

In a symmetry-breaking with a high D-band peak, a charge carrier must be excited and inelastically scattered by a phonon (the D-band denoting disorder-induced). As the amount of disorder in graphene increases, the Raman intensity increases for the different disorder peaks D (1350 cm$^{-1}$), which scatters from K to K′ (internally); D′ (1620 cm$^{-1}$), which scatters from K to K (intravalley) [67]. It has been reported that the ratio of D, $I_D/I_G$ and $I_D/I_{D'}$ intensity can be used to discriminate the disorder, sp$^3$ and vacancy-type defects in graphene-based devices [50], [68]. Fig. 6(a) shows a forward breakdown of the GNRs device under high electrical stress. The SEM images and Raman spectra in Fig. 6(b) and Fig 6(c) show high D, D′, and large $I_D/I_G$>2 peak ratio, revealing a higher defect density, breaking the C-C symmetry, and more phonon dispersion after electrical stress breakdown. The GNRs that suffer higher charge crowding is less efficient for heat management, leading to phonon scatting and potentially damaging the Schottky junction, as presented in Fig. 6(b) and Fig. 6(c). The GNRs have sharp and distinct uniform boundaries before electrical stress (Fig. 6(d-f)). The sharp GNRs changed into non-uniform and irregular termination at edges resulting from the disruption due to high electrical stress shown in Raman mapping (Fig. 6(g-i)).

Gr/Si Schottky devices broke down due to graphene rupturing at edge contacts rather than the active device area. So, the Gr/Si junction is unaffected and intact even after graphene is burnt on contact edges. We fabricated a multi-electrode device with the same Gr/Si junction for a broader investigation. The I-V curves of multiple electrodes are shown in Fig.7(a). A 3D schematic of multiple electrodes is shown in an inset of Fig. 7(a). When the device loses its graphene contact with one electrode, the other electrodes can act as an additional electrode, enhancing its usability even after multiple breakdowns. As illustrated in Fig. 7(b), when E-1 was undergoing high electrical stress, the neighboring electrode-2 (E-2) remained unaffected. When electrical stress was applied to E-2, the device was still functional and failed at 103 V. The above-stated technique, as Gr/Si Schottky junction, is thermally stable under ESD conditions and can use a single junction multiple times.

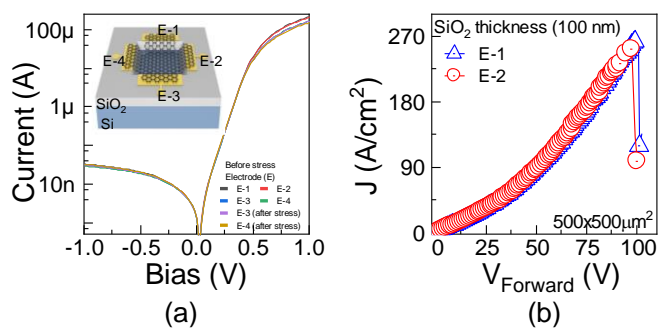

**Fig. 7.** (a) I-V characterization of multiple electrodes of Gr/Si junction (Inset: 3D schematic of multiple electrodes). (b) Gr/Si junction current density is a function of the forward voltage ($V_F$) at electrodes E-1 and E-2 after a breakdown.

In the light of the present work, it is shown that the Gr/Si Schottky diode is a suitable candidate for being utilized in high frequency and charge dynamics due to its proven robustness. The fast response and absorption of ESD spiking lead us to new applications to protect spike-sensitive devices

## V. CONCLUSION

The 2D-3D heterostructures are essential to electrical and optoelectronic study as providing a fundamental approach to exploring charge transport. The challenges of 2D-3D heterostructures lie in ensuring their robustness, highly durable, and reliability, which demands extensive study. We have used transmission line pulse and microscopic techniques to explore the electrical charge dynamic at 2D-3D contact. Our data indicate that charge crowding significantly limits the device performance. The role of current crowding will reduce, and charge transport will enhance by improving the heat mitigation. It has important facts for the design of graphene-based devices, as the graphene ribbon device appears to be more defective and prone to disruptions under high electrical stress. In contrast, the Gr/Si Schottky junction (with graphene sheet) is structurally stable and robust without sacrificing its junction performance. Benefitting its broadband photodetection in parallel, it has excellent potential for applications where high current carrying capability and avalanche photodetection at large voltage biases. Our work also draws attention to the significance of engineering the 2D-3D interface, which often is the bottleneck preventing the nanoscale 2D-3D device far from reaching its ideal performance.

REFERENCES

[1] A. A. Balandin *et al.*, "Superior Thermal Conductivity of Single-Layer Graphene," *Nano Lett.*, vol. 8, no. 3, pp. 902–907, 2008.
[2] Y. Liu *et al.*, "Approaching the Schottky–Mott limit in van der Waals metal–semiconductor junctions," *Nature*, vol. 557, pp. 696–700, 2018.
[3] K. I. Bolotin *et al.*, "Ultrahigh electron mobility in suspended graphene," *Solid State Commun.*, vol. 146, pp. 351–355, 2008.
[4] E. Kim, N. Jain, R. Jacobs-Gedrim, Y. Xu, and B. Yu, "Exploring carrier transport phenomena in a CVD-assembled graphene FET on hexagonal boron nitride," *Nanotechnology*, vol. 23, no. 12, p. 125706, 2012, doi: 10.1088/0957-4484/23/12/125706.
[5] I. W. Frank, D. M. Tanenbaum, A. M. Van Der Zande, P. L. Mceuen, and N. S. Processing, "Mechanical properties of suspended graphene sheets," *J. Vac. Sci. Technol. B*, vol. 25, pp. 2558–2561, 2007, doi: 10.1116/1.2789446.
[6] A. Ali *et al.*, "High-performance, flexible graphene/ultra-thin silicon ultra-violet image sensor," *IEDM*, pp. 8.6.1-8.6.4, 2017.
[7] K. V Zakharchenko, M. I. Katsnelson, and A. Fasolino, "Melting temperature of graphene," *Phys. Rev. B*, vol. 91, p. 45415, 2015.
[8] S. Du *et al.*, "A Broadband Fluorographene Photodetector," *Adv. Mater*, vol. 29, no. 22, p. 1700463, 2017, doi: 10.1002/adma.201700463.
[9] R. Hao *et al.*, "Improved Slow Light Capacity In Graphene-based Waveguide," *Sci. Rep.*, vol. 5, no. 1, p. 15335, 2015, doi: 10.1038/srep15335.
[10] K. L. Grosse, M. Bae, F. Lian, E. Pop, and W. P. King, "Nanoscale Joule heating , Peltier cooling and current crowding at graphene – metal contacts," *Nat. Nanotechnol.*, vol. 6, pp. 287–290, 2011.
[11] A. Bar-Cohen and P. Wang, "Thermal management of on-chip hot spot," *J. Heat Transfer*, vol. 134, no. 5, p. 051017, 2012, doi: 10.1115/1.4005708.
[12] S. W. Liang, Y. W. Chang, and C. Chen, "Relieving hot-spot temperature and current crowding effects during electromigration in solder bumps by using Cu columns," *J. Electron. Mater.*, vol. 36, no. 10, pp. 1348–1354, 2007, doi: 10.1007/s11664-007-0232-3.
[13] S. Banerjee, J. Luginsland, and P. Zhang, "Interface Engineering of Electrical Contacts," *Phys. Rev. Appl.*, vol. 15, no. 6, p. 64048, 2021, doi: 10.1103/PhysRevApplied.15.064048.
[14] A. Allain, J. Kang, K. Banerjee, and A. Kis, "Electrical contacts to two-dimensional semiconductors," *Nat. Mater.*, vol. 14, pp. 1195–1205, 2015.
[15] M. Pedram, "Power minimization in 1c design:

principles and applications," *ACM Trans. Des. Autom. Electron. Syst.*, vol. 1, no. 1, pp. 3–56, 1996, doi: 10.1145/225871.225877.

[16] C. N. Liao and K. C. Chen, "Current Crowding Effect on Thermal Characteristics of Ni/Doped-Si Contacts," *IEEE Electron Device Lett.*, vol. 24, no. 10, pp. 637–639, 2003, doi: 10.1109/LED.2003.817875.

[17] S. Min Song, T. Yong Kim, O. Jae Sul, W. Cheol Shin, and B. Jin Cho, "Improvement of graphene-metal contact resistance by introducing edge contacts at graphene under metal," *Appl. Phys. Lett.*, vol. 104, no. 18, p. 183506, 2014, doi: 10.1063/1.4875709.

[18] P. Karnatak, T. P. Sai, S. Goswami, S. Ghatak, S. Kaushal, and A. Ghosh, "Current crowding mediated large contact noise in graphene field-effect transistors," *Nat. Commun.*, vol. 7, p. 13703, 2016, doi: 10.1038/ncomms13703.

[19] P. Zhang, Y. Y. Lau, and R. M. Gilgenbach, "Analysis of current crowding in thin film contacts from exact field solution," *J. Phys. D Appl. Phys.*, vol. 48, no. 47, p. 475501, 2015, doi: 10.1088/0022-3727/48/47/475501.

[20] H. Murrmann and D. Widmann, "Current crowding on metal contacts to planar devices," *IEEE Trans. Electron Devices*, vol. 16, no. 12, pp. 1022–1024, 1969.

[21] S. Banerjee, P. Y. Wong, and P. Zhang, "Contact resistance and current crowding in tunneling type circular nano-contacts," *J. Phys. D. Appl. Phys.*, vol. 53, no. 35, p. 355301, 2020.

[22] P. Yang, S. Banerjee, W. Kuang, Y. Ding, Q. Ma, and P. Zhang, "Current crowding and spreading resistance of electrical contacts with irregular contact edges," *J. Phys. D Appl. Phys*, vol. 53, no. 48, p. 485303, 2020, doi: 10.1088/1361-6463/abadc3.

[23] H. H. Berger, "Contact Resistance and Contact Resistivity," *J. Electrochem. Soc.*, vol. 119, no. 4, p. 507, 1972, doi: 10.1149/1.2404240.

[24] I. Ratković, N. Bežanić, O. S. Ünsal, A. Cristal, and V. Milutinović, "An Overview of Architecture-Level Power- and Energy-Efficient Design Techniques," *Adv. Comput.*, vol. 98, pp. 1–57, 2015, doi: 10.1016/bs.adcom.2015.04.001.

[25] Q. Wang, X. Tao, L. Yang, and Y. Gu, "Current crowding in two-dimensional black-phosphorus field-effect transistors," *Appl. Phys. Lett.*, vol. 108, no. 10, p. 103109, 2016, doi: 10.1063/1.4943655.

[26] H. Yuan et al., "Field effects of current crowding in metal-MoS2 contacts," *Appl. Phys. Lett.*, vol. 108, p. 103505, 2016, doi: 10.1063/1.4942409.

[27] K. Nagashio, T. Nishimura, K. Kita, and A. Toriumi, "Contact resistivity and current flow path at metal/graphene contact," *Appl. Phys. Lett.*, vol. 97, no. 14, p. 143514, 2010, doi: 10.1063/1.3491804.

[28] C. W. T. Chiang and Y. C. S. Chang, "Reducing the Current Crowding Effect on Nitride- Based Light-Emitting Diodes Using Modulated P- Extension Electrode Thickness," *Jpn. J. Appl. Phys.*, vol. 52, no. 1S, p. 01AG05, 2013.

[29] N. Shamir and D. Ritter, "Reducing the current crowding effect in bipolar transistors by tunnel diode emitter design," *Solid. State. Electron.*, vol. 46, pp. 127–130, 2003.

[30] A. Phys, "Reduction of on-resistance and current crowding in quasi-vertical GaN power diodes," *Appl. Phys. Lett.*, vol. 111, no. June, p. 163506, 2017, doi: 10.1063/1.4989599.

[31] M. Massicotte et al., "Photo-thermionic effect in vertical graphene heterostructures," *Nat. Commun.*, vol. 7, p. 12174, 2016, doi: 10.1038/ncomms12174.

[32] A. A. Balandin, "Low-frequency 1 / f noise in graphene devices," *Nat. Nanotechnol.*, vol. 8, no. August, pp. 549–555, 2013, doi: 10.1038/nnano.2013.144.

[33] Z. Chen et al., "High Responsivity, Broadband, and Fast Graphene/Silicon Photodetector in Photoconductor Mode," *Adv. Opt. Mater.*, vol. 3, no. 9, pp. 1207–1214, 2015, doi: 10.1002/adom.201500127.

[34] C. H. Liu, Y. C. Chang, T. B. Norris, and Z. Zhong, "Graphene photodetectors with ultra-broadband and high responsivity at room temperature," *Nat. Nanotechnol.*, vol. 9, no. 4, pp. 273–278, 2014, doi: 10.1038/nnano.2014.31.

[35] F. and A. D. B. Giubileo, "The role of contact resistance in graphene field-effect devices Filippo Giubileo," *Prog. Surf. Sci.*, vol. 92, pp. 143–175, 2017.

[36] G. Giovannetti, P. A. Khomyakov, G. Brocks, V. M. Karpan, J. Van Den Brink, and P. J. Kelly, "Doping Graphene with Metal Contacts," *Phys. Rev. Lett.*, vol. 101, no. 2, p. 26803, 2008, doi: 10.1103/PhysRevLett.101.026803.

[37] H. Li et al., "On the Electrostatic Discharge Robustness of Graphene," *IEEE Trans. Electron Devices*, vol. 61, no. 6, pp. 1920–1928, 2014.

[38] Z. Xu and M. J. Buehler, "Heat dissipation at a graphene – substrate interface," *J. Phys. Condens. Matter*, vol. 24, p. 475305, 2012.

[39] A. De Iacovo, L. Colace, G. Assanto, L. Maiolo, and A. Pecora, "Extraction of Schottky barrier parameters for metal-semiconductor junctions on high resistivity inhomogeneous, semiconductors," *IEEE Trans. Electron Devices*, vol. 62, no. 2, pp. 465–470, 2015, doi: 10.1109/TED.2014.2378015.

[40] J. Xiaoting et al., "Controlled Formation of Sharp Zigzag and Armchair Edges in Graphitic Nanoribbons," *Science (80-. ).*, vol. 323, no. 5922, pp. 1701–1706, 2009.

[41] Y. W. Tan et al., "Graphene at the Edge: Stability and Dynamics," *Scienc*, vol. 666, no. March, pp. 1705–1709, 2009.

[42] C. Durkan and Z. Xiao, "On the failure of graphene devices by Joule heating under current stressing conditions," *Appl. Phys. Lett.*, vol. 107, no. 24, p. 243505, 2015, doi: 10.1063/1.4936993.

[43] P. Shukla, A. Subramani, and A. Sen, "Grain-Boundary Effects on the Charge Transport Behavior of Quasi- 2D Graphene/PVDF for Electrostatic Control of Power Dissipation in GFETs," *J. Phys. Chem. C.*, vol. 125, p. 10441−10450, 2021, doi: 10.1021/acs.jpcc.1c00940.

[44] V. A. Vashchenko and V. F. Sinkevitch, *Physical Limitations of Semiconductor Devices*. Springer, Boston, MA, 2008.

[45] A. D. Liao, A. Behnam, V. E. Dorgan, Z. Li, and E. Pop, "Reliability, failure, and fundamental limits of graphene and carbon nanotube interconnects," *Tech. Dig. - Int. Electron Devices Meet. IEDM*, pp. 15.1.1-15.1.4, 2013.

[46] X. Li *et al.*, "Strong substrate effects of Joule heating in graphene electronics Strong substrate effects of Joule heating in graphene electronics," *Appl. Phys. Lett.*, vol. 99, no. 23, p. 233114, 2011, doi: 10.1063/1.3668113.

[47] C. Zhu *et al.*, "Investigation of microwave and noise properties of InAlN / GaN HFETs after electrical stress : Role of surface effects," *Proc. SPIE*, vol. 8625, no. 86252H, pp. 389–395, 2013.

[48] M. S. Dresselhaus, A. Jorio, A. G. S. Filho, and R. Saito, "Defect characterization in graphene and carbon nanotubes using Raman spectroscopy," *Phil. Trans. R. Soc. A*, vol. 368, pp. 5355–5377, 2010.

[49] A. Zandiatashbar *et al.*, "Effect of defects on the intrinsic strength and stiffness of graphene," *Nat. Commun.*, vol. 5, no. 1, p. 3186, 2014, doi: 10.1038/ncomms4186.

[50] A. Eckmann *et al.*, "Probing the nature of defects in graphene by Raman spectroscopy," *Nano Lett.*, vol. 12, no. 8, pp. 3925–3930, 2012, doi: 10.1021/nl300901a.

[51] A. C. Ferrari, "Raman spectroscopy of graphene and graphite : Disorder , electron – phonon coupling , doping and nonadiabatic effects," *Solid State Commun.*, vol. 143, no. 1, pp. 47–57, 2007, doi: 10.1016/j.ssc.2007.03.052.

[52] P. T. Araujo, M. Terrones, and M. S. Dresselhaus, "Defects and impurities in graphene-like materials," *Mater. Today*, vol. 15, no. 3, pp. 98–109, 2012, doi: 10.1016/S1369-7021(12)70045-7.

[53] J. Hwang, C. Kuo, and L. Chen, "Correlating defect density with carrier mobility in large-scaled graphene films: Raman spectral signatures for the estimation of defect density," *Nanotechnology*, vol. 21, no. 46, p. 465705, 2010, doi: 10.1088/0957-4484/21/46/465705.

[54] A. H. Castro Neto, F. Guinea, N. M. R. Peres, K. S. Novoselov, and A. K. Geim, "The electronic properties of graphene," *Rev. Mod. Phys.*, vol. 81, no. 1, pp. 109–162, Jan. 2009, doi: 10.1103/RevModPhys.81.109.

[55] K. Brenner, T. J. Beck, and J. D. Meindl, "Breakdown current density of graphene nanoribbons Breakdown current density of graphene nanoribbons," *Appl. Phys. Lett.*, vol. 94, no. 24, p. 243114, 2009, doi: 10.1063/1.3147183.

[56] H. Li, S. Xiao, H. Yu, Y. Xue, and J. Yang, "A review of graphene-based films for heat dissipation," *New Carbon Mater.*, vol. 36, no. 5, pp. 897–908, 2021, doi: 10.1016/S1872-5805(21)60092-6.

[57] A. H. C. Neto, "The electronic properties of graphene," *Rev. Mod. Phys.*, vol. 81, no. 1, pp. 109–162, 2009, doi: 10.1103/RevModPhys.81.109.

[58] F. Tuinstra and J. L. Koenig, "Raman Spectrum of Graphite Raman Spectrum of Graphite," *J. Chem. Phys.*, vol. 53, pp. 1126–1130, 1970, doi: 10.1063/1.1674108.

[59] M. S. Islam, S. Tanaka, and A. Hashimoto, "Effect of vacancy defects on phonon properties of hydrogen passivated graphene nanoribbons," *Carbon.*, vol. 80, pp. 146–154, 2014.

[60] Z. Xie, Y. Zhang, L. Zhang, and D. Fan, "Effect of topological line defects on electron-derived thermal transport in zigzag graphene nanoribbons," *Carbon.*, vol. 113, pp. 292–298, 2017.

[61] E. R. Mucciolo, A. H. Castro Neto, and C. H. Lewenkopf, "Conductance quantization and transport gaps in disordered graphene nanoribbons," *Phys. Rev. B*, vol. 79, no. 7, p. 75407, 2009, doi: 10.1103/PhysRevB.79.075407.

[62] T. C. Li and S. Lu, "Quantum conductance of graphene nanoribbons with edge defects," *Phys. Rev. B.*, vol. 77, no. 8, p. 85408, 2008, doi: 10.1103/PhysRevB.77.085408.

[63] F. Banhart, J. Kotakoski, and A. V Krasheninnikov, "Structural Defects in Graphene," *ACS Nano*, vol. 5, no. 1, pp. 26–41, 2011.

[64] K. Rallis *et al.*, "Electronic Properties of Graphene Nanoribbons With Defects," *IEEE Trans. Nanotechnol.*, vol. 20, pp. 151–160, 2021.

[65] K. Nakada, M. Fujita, G. Dresselhaus, and M. S. Dresselhaus, "Edge state in graphene ribbons : Nanometer size effect and edge shape dependence," *Phys.Rev.B*, vol. 54, no. 24, pp. 954–961, 1996.

[66] W. Zhang *et al.*, "The electro-mechanical responses of suspended graphene ribbons for electrostatic discharge applications," *Appl. Phys. Lett.*, vol. 108, no. 15, p. 153103, 2016, doi: 10.1063/1.4946007.

[67] R. Saito *et al.*, "Advances in Physics Raman spectroscopy of graphene and carbon nanotubes," *Adv. Phys.*, vol. 60, no. 3, pp. 413–550, 2011, doi: 10.1080/00018732.2011.582251.

[68] E. H. M. Ferreira *et al.*, "Evolution of the Raman spectra from single- , few- , and many-layer graphene with increasing disorder," *Phys. Rev. B*, vol. 82, p. 125429, 2010, doi: 10.1103/PhysRevB.82.125429.